# Topological pumping in photonic systems

Solange V. Silva[1], David E. Fernandes[1], Tiago A. Morgado[1], Mário G. Silveirinha[2*]

*[1]Instituto de Telecomunicações and Department of Electrical Engineering, University of Coimbra, 3030-290 Coimbra, Portugal*

*[2]University of Lisbon, Instituto Superior Técnico, Avenida Rovisco Pais, 1, 1049-001 Lisboa, Portugal*

*E-mail:* **solange@co.it.pt**, **dfernandes@co.it.pt, tiago.morgado@co.it.pt**, **mario.silveirinha@co.it.pt**

## Abstract

The topology of typical Chern insulators is rooted in the periodicity of the system along two directions of real-space. In this article, we depart from this standard concept and demonstrate that a generic non-Hermitian photonic waveguide periodic along a single direction of real space can be regarded as a sub-component of an extended system with a synthetic dimension and with a nontrivial Chern topology. In particular, we show that the number of bands below a band-gap of a generic waveguide determines the gap Chern number of the extended system. It is theoretically and numerically demonstrated that in real-space the gap Chern number gives the number of gapless Tamm state branches localized at the system boundary, when its geometry is continuously displaced by one lattice period. In the non-Hermitian case, the Tamm states connect different bands in the complex plane.

[*] To whom correspondence should be addressed: E-mail: mario.silveirinha@co.it.pt

# I. Introduction

Topology has recently emerged as a new tool to characterize global properties of physical systems, e.g., physical responses that are robust to perturbations of the system parameters [1-11]. There are different classes of topological platforms. Usually a nontrivial topology is rooted in some particular symmetry or combination of symmetries of the system, e.g., invariance under discrete translations, time-reversal, parity, etc. For systems with a Chern-type classification the topological analysis relies on the spectrum of some family of Hermitian operators $\hat{H}_{\mathbf{q}}$ parameterized by a two-component label $\mathbf{q}=(q_1,q_2)$ [12-13]. The Hermitian property is not essential [13-21]. Provided the two-parameter space is a closed surface with no boundary and $\hat{H}_{\mathbf{q}}$ varies smoothly with $\mathbf{q}$, then it is possible to assign a topological number $\mathcal{C}_{\text{gap}}$ to the spectral band-gaps. This result is known as the Chern theorem. The number $\mathcal{C}_{\text{gap}}$ is an integer and its value is insensitive to perturbations of $\hat{H}_{\mathbf{q}}$ that do not close a band-gap.

In most studies so far, the topological properties are inherited from the periodicity of the system along two-directions of real-space and $\mathbf{q}$ is identified with a Bloch wave vector. The corresponding two-parameter space is a Brillouin zone, which is effectively a closed surface with no boundary (a torus) due to its cyclic nature. Here, we extend the Chern classification to generic 1D-type photonic platforms, e.g., an arbitrary waveguide that supports propagation along a fixed direction of space. It is shown that any 1D-type periodic system can be regarded as a topological system with a synthetic dimension, and we theoretically and numerically demonstrate that the number of photonic bands below the gap is identical to the gap Chern number. Furthermore, it is shown that edge states in the extended system with a synthetic dimension are mapped into Tamm states in the real-space, i.e., to excitations localized at the end of the 1D photonic guide. In particular, it is demonstrated that the bulk-edge correspondence implies that the number of gapless Tamm states created when the geometry of the 1D-periodic system is continuously displaced by one spatial period is determined by a difference of topological numbers. Finally, we demonstrate that the outlined ideas can be extended to non-Hermitian waveguides.

It should be noted that previous works [22-24] predicted topological light-trapping on dislocations, but using mechanisms different from ours. Furthermore, topological classifications of particular sub-classes of 1D-systems have been previously developed by other authors using Zak phases, winding numbers, and related concepts [24-28].

## II. Topological band count

We consider a generic platform that is formed by a 1D real-space periodic system, which we shall designate as "waveguide". The "waveguide" can be visualized as some periodic (possibly three-dimensional (3D)) structure that only allows propagation (waveguiding) along some fixed direction, let us say the $x$-direction. For example, it can be a hollow metallic structure, with the metal walls invariant to translations along the $x$-axis, and with the guide periodically loaded with dielectric inclusions $\varepsilon(x,y,z)=\varepsilon(x+a,y,z)$; here $a$ is the lattice period. For simplicity, in most examples we shall take the "waveguide" as a genuinely 1D photonic crystal formed by a periodic stack of dielectric slabs ($\varepsilon(x)=\varepsilon(x+a)$) and restrict our attention to propagation along the $x$ axis. However, it is underlined that it can be fully three-dimensional.

We admit that the wave propagation in the structure is determined by some operator $\hat{H}(\mathbf{r},-i\nabla)$ such that the time evolution of the system state vector $\psi$, e.g., the electromagnetic field, is described by Schrödinger-type dynamics $i\partial_t\psi=\hat{H}\psi$. The time evolution of any (eventually dispersive) electromagnetic platform can always be expressed in such a manner [9, 12, 29, 30]. For convenience, we designate $\hat{H}$ as the Hamiltonian.

Due to the periodicity along the $x$-direction, the eigenstates are Bloch waves labeled by a Bloch wave number $q_x$. The corresponding envelopes $u_{q_x}$ (defined such that $\psi_{q_x} = u_{q_x} e^{iq_x x}$) satisfy $\hat{H}(x, -i\partial_x + q_x) u_{q_x} = \omega_{q_x} u_{q_x}$ with $\omega_{q_x}$ the eigenfrequencies. Note that $u_{q_x}$ can be a multi-component vector. The parameters $y, z, -i\partial_y, -i\partial_z$ are omitted from now on in the argument of the operator $\hat{H}$ as they are not relevant for the discussion.

Let us now add a second label ($q_s$) to the Hamiltonian related to a translation in space $x \to x - x_0$:

$$\hat{H}_{\mathbf{q}} \equiv \hat{H}_{q_x, q_s} = \hat{H}\left(x - x_0(q_s), -i\partial_x + q_x\right). \tag{1}$$

The coordinate shift $x_0$ is parameterized by $q_s$. In section III, it will be shown that $q_s$ may be understood as a "momentum" determined by a synthetic dimension. It is assumed that $x_0(q_s)$ is a smooth function and that $x_0(q_s + 2\pi) - x_0(q_s) = Na$, with $a$ the spatial period of the waveguide and $N$ some integer number. Since $\hat{H}(x, -i\partial_x + q_x) = \hat{H}(x - a, -i\partial_x + q_x)$, it follows that $\hat{H}_{q_x, q_s}$ is a periodic function of $q_s$ with period $2\pi$. In a full cycle, as $q_s$ varies from $q_s = -\pi$ to $q_s = \pi$, the waveguide is displaced by $N$ complete spatial periods towards the $+x$-direction.

Since the spectrum of $\hat{H}_{q_x, q_s}$ is cyclic in both $q_x$ and $q_s$ one can characterize its topological phases. To this end, consider a generic band of eigenfunctions ($\psi_{q_x}(x)$) of the "waveguide": $\hat{H}(x, -i\partial_x)\psi_{q_x}(x) = \omega_{q_x}\psi_{q_x}(x)$. Then, it is obvious that $\hat{H}_{\mathbf{q}} u_{\mathbf{q}} = \omega_{\mathbf{q}} u_{\mathbf{q}}$ with $\omega_{\mathbf{q}} = \omega_{q_x}$, $\mathbf{q} = (q_x, q_s)$ and with envelope given by

$$u_{\mathbf{q}}(x) = \psi_{q_x}\left(x - x_0(q_s)\right) e^{-iq_x x} = u_{q_x}\left(x - x_0(q_s)\right) e^{-iq_x x_0(q_s)}. \tag{2}$$

Clearly, the eigenvalues of $\hat{H}_{\mathbf{q}}$ are independent of $q_s$ and thereby the band-gaps of $\hat{H}_{\mathbf{q}}$ are the same as the band-gaps of the “waveguide”. In other words, a translation in space does not alter the band structure.

In the following it is assumed for simplicity that $\hat{H}$ is a Hermitian operator. The generalization of the analysis to the non-Hermitian case is reported in Appendix A. The Bloch eigenmodes $\psi_{q_x}$ of $\hat{H}$ can be taken as smooth periodic functions of $q_x$ in the 1D-Brillouin zone $-\pi / a \leq q_x \leq \pi / a$. The Chern number $\mathcal{C}$ associated with a given band of $\hat{H}_{\mathbf{q}}$ can be found in a standard way from the Berry potential $\mathcal{A}_{\mathbf{q}} = i\langle u_{\mathbf{q}} | \partial_{\mathbf{q}} u_{\mathbf{q}} \rangle$ using $\mathcal{C} = \frac{1}{2\pi} \int_{-\pi/a}^{\pi/a} dq_1 \int_{-\pi}^{\pi} dq_2 \left( \frac{\partial \mathcal{A}_{2,\mathbf{q}}}{\partial q_1} - \frac{\partial \mathcal{A}_{1,\mathbf{q}}}{\partial q_2} \right)$ with $(q_1, q_2) \equiv (q_x, q_s)$. The eigenfunctions are normalized as $\langle u_{\mathbf{q}} | u_{\mathbf{q}} \rangle = 1$ with $\langle . | . \rangle$ the canonical inner product. Since the Berry potential is a smooth function in the interior of the integration domain, from the Stokes theorem the Chern number is:

$$
\begin{aligned}
\mathcal{C} &= \frac{1}{2\pi} \int_{-\pi/a}^{\pi/a} dq_x \left( \mathcal{A}_{1,\mathbf{q}}\big|_{q_s=-\pi} - \mathcal{A}_{1,\mathbf{q}}\big|_{q_s=+\pi} \right) \\
&+ \frac{1}{2\pi} \int_{-\pi}^{\pi} dq_s \left( \mathcal{A}_{2,\mathbf{q}}\big|_{q_x=\pi/a} - \mathcal{A}_{2,\mathbf{q}}\big|_{q_x=-\pi/a} \right).
\end{aligned} \tag{3}
$$

Using $u_{\mathbf{q}}(x) = \psi_{q_x}(x - x_0(q_s)) e^{-iq_x x}$ one finds that $\mathcal{A}_{2,\mathbf{q}} = i\langle \psi_{\mathbf{q}} | \partial_{q_s} \psi_{\mathbf{q}} \rangle$ with $\psi_{\mathbf{q}} \equiv \psi_{q_x}(x - x_0(q_s))$. Noting that $i\langle \psi_{\mathbf{q}} | \partial_{q_s} \psi_{\mathbf{q}} \rangle$ is a periodic function of $q_x$, it follows that the second integral in the right-hand side of Eq. (3) vanishes. On the other hand, using $u_{\mathbf{q}}(x) = u_{q_x}(x - x_0(q_s)) e^{-iq_x x_0(q_s)}$, we get

$$
\mathcal{A}_{1,\mathbf{q}} = i\left\langle u_{q_x}(x - x_0(q_s)) \,|\, \partial_{q_x}\left[ u_{q_x}(x - x_0(q_s)) \right] \right\rangle + x_0(q_s). \tag{4}
$$

We used $\langle u_{\mathbf{q}} | u_{\mathbf{q}} \rangle = 1$ and the periodicity of the envelope in *x*. The first term in the right-hand side of Eq. (4) is a periodic function of $q_s$ because of the periodicity of the envelope in *x*. Thus, it does not contribute to the first integral in Eq. (3). Taking this into account, we obtain the key result:

$$\begin{aligned} \mathcal{C} &= \frac{1}{2\pi} \int_{-\pi/a}^{\pi/a} dq_x \left( x_0\left(-\pi\right) - x_0\left(+\pi\right) \right) \\ &= -\frac{1}{a}\left[ x_0\left(\pi\right) - x_0\left(-\pi\right) \right] = -N. \end{aligned} \tag{5}$$

We used $x_0\left(q_s + 2\pi\right) - x_0\left(q_s\right) = Na$ in the last identity. The above formula proves that each photonic band of $\hat{H}_{\mathbf{q}}$ has a topological charge of "–*N*", i.e., identical to minus the number of displaced unit cells towards +*x*. Note that $\mathcal{C}$ is an integer. In particular, the gap Chern number of a given band-gap is identical to the number of bands ($n_{\text{bands}}$) below the gap multiplied by the number of shifted cells (*N*) in one $q_s$-cycle:

$$\mathcal{C}_{\text{gap}} = -n_{\text{bands}} \times N\,. \tag{6}$$

This means that the number of photonic bands below the gap of a generic 1D-type photonic crystal can be understood as a (topological) Chern number. The topological charge of each band is acquired from the translational shift suffered by the waveguide. In other words, a translation in space effectively "pumps" topological charge into the system described by $\hat{H}_{\mathbf{q}}$. This property and its consequences are discussed in Section III.

# III. The synthetic dimension and bulk-edge correspondence

## *A. The synthetic dimension*

Next, it is shown that $\hat{H}_{\mathbf{q}}$ can be regarded as the momentum-space operator of a system that consists of the original 1D-type waveguide (which as previously mentioned can be embedded in a three-dimensional space) with an additional synthetic dimension. Systems with synthetic dimensions were recently discussed in the literature to emulate physical and topological phenomena in higher dimensions (see e.g., [31-33]).

Consider a generic family of operators $\hat{H}_{\mathcal{K}}\left(x,-i\partial_x\right)$ periodic both in $\mathcal{K}$ and $x$: $\hat{H}_{\mathcal{K}}=\hat{H}_{\mathcal{K}+2\pi}$ and $\hat{H}_{\mathcal{K}}\left(x,-i\partial_x\right)=\hat{H}_{\mathcal{K}}\left(x+a,-i\partial_x\right)$. The operator $\hat{H}_{\mathcal{K}}\left(x,-i\partial_x\right)$ may also depend on other space coordinates ($y$, $z$, etc) and space derivatives, but since they are not relevant for the analysis they are omitted. We introduce a matrix operator ($\hat{\mathcal{H}}_e$) that acts on a column state vector of the form $\Psi=\left[\psi_m\left(x\right)\right]=[\ldots\ \ \psi_{-1}\ \ \psi_0\ \ \psi_1\ \ \ldots]^T$, $m$=0,±1, ±2,…, through a convolution:

$$\Psi\rightarrow\hat{\mathcal{H}}_e\Psi=\left[\left(\hat{\mathcal{H}}_e\Psi\right)_n\right]\quad\text{where}\quad\left(\hat{\mathcal{H}}_e\Psi\right)_n=\sum_m\hat{H}_{n-m}\left(x,-i\partial_x\right)\psi_m\left(x\right),\tag{7}$$

with $n$=0,±1, ±2,… . The matrix elements of $\hat{\mathcal{H}}_e$ are defined as:

$$\hat{H}_m\left(x,-i\partial_x\right)=\frac{1}{2\pi}\int_0^{2\pi}d\mathcal{K}\,\hat{H}_{\mathcal{K}}\left(x,-i\partial_x\right)e^{i\mathcal{K}m},\qquad m=0,\pm1,\pm2,\ldots\tag{8}$$

The state vector $\Psi=\left[\psi_m\left(x\right)\right]$ has two space-type coordinates: $x$ which corresponds to a continuous real-space coordinate, and $m$ which corresponds to a discrete (lattice) coordinate. The coordinate $m$ determines the synthetic dimension. The Bloch eigenfunctions are characterized by a state vector of the form $\Psi=\left[\psi_m\left(x\right)\right]$ with $\psi_m\left(x\right)=u_{k,\mathcal{K}}\left(x\right)e^{ikx}e^{im\mathcal{K}}$ and satisfy $\hat{\mathcal{H}}_e\Psi=\omega_{k,\mathcal{K}}\Psi$, where $\left(k,\mathcal{K}\right)$ is the two-dimensional Bloch wave vector ($-\pi/a\leq k\leq\pi/a$ and $-\pi\leq\mathcal{K}\leq\pi$). Substituting $\Psi=\left[u_{k,\mathcal{K}}\left(x\right)e^{ikx}e^{im\mathcal{K}}\right]$ into Eq. (7) and using the Fourier synthesis relation $\hat{H}_{\mathcal{K}}\left(x,-i\partial_x\right)=\sum_n\hat{H}_n\left(x,-i\partial_x\right)e^{-in\mathcal{K}}$, it is found that the secular equation $\hat{\mathcal{H}}_e\Psi=\omega_{k,\mathcal{K}}\Psi$ reduces to:

$$\hat{H}_{\mathcal{K}}\left(x,-i\partial_x+k\right)u_{k,\mathcal{K}}=\omega_{k,\mathcal{K}}u_{k,\mathcal{K}}\,. \tag{9}$$

Thus, the operator $\hat{H}_{\mathcal{K}}\left(x,-i\partial_x+k\right)$ is the momentum-space version of $\hat{\mathcal{H}}_e$.

The previous theory can be readily applied to the family of operators $\hat{H}_{\mathbf{q}}=\hat{H}\left(x-x_0\left(q_s\right),-i\partial_x+q_x\right)$ considered in Sect. II, with the obvious correspondence $\left(q_x,q_s\right)\leftrightarrow\left(k,\mathcal{K}\right)$. In particular, $\hat{H}_{\mathbf{q}}$ is the momentum-space version of some operator $\hat{\mathcal{H}}_e$ defined on an extended-space determined by the continuous coordinate $x$ (which varies in the "real-space") and by the discrete coordinate $m$ (which varies along the synthetic lattice-type dimension). This property is important as it guarantees that the gap-Chern number can be linked to the number of edge-states through a bulk-edge correspondence [13, 34-38].

## *B. The bulk-edge correspondence*

The bulk-edge correspondence establishes a precise relation between the gap Chern numbers of two topological materials and the net number of unidirectional edge states [34-40]. Thus, the Chern invariants of the operator $\hat{\mathcal{H}}_e$ determine a bulk-edge correspondence in the extended space with a synthetic dimension. An obvious question is: what are the consequences of the bulk-edge correspondence in real-space?

To address this point, consider two 1D-type periodic "waveguides", described by the (real-space) Hamiltonians $\hat{H}_1$ and $\hat{H}_2$, respectively. Suppose that the waveguides have a common band-gap. Furthermore, let us add a synthetic (discrete) dimension to each waveguide, such that the extended-space Hamiltonians are $\hat{\mathcal{H}}_{e1}$ and $\hat{\mathcal{H}}_{e2}$, with each of them described by a momentum-space Hamiltonian of the form $\hat{H}_{\mathbf{q},i}=\hat{H}_i\left(x-x_0^{(i)}\left(q_s\right),-i\partial_x+q_x\right)$, $i$=1,2. For definiteness, we take $x_0^{(i)}\left(q_s\right)\equiv N_i\Delta$, with $N_i$ an integer and $\Delta=\frac{q_s}{2\pi}a$. Then, from Eq. (6) the gap Chern number difference in a common gap is:

$$\begin{aligned}\delta\mathcal{C}_{\text{gap}} &\equiv \mathcal{C}_{\text{gap},1} - \mathcal{C}_{\text{gap},2} \\ &= n_{\text{bands},2} \times N_2 - n_{\text{bands},1} \times N_1\end{aligned}. \tag{10}$$

Here, $n_{\text{bands},i}$ is the number of bands below the gap for the *i*-th waveguide. In particular, when the number of shifted cells in both waveguides is $N_2 = N_1 = 1$, the gap Chern number difference is given by the difference of the number of bands below the gap, which thereby is a topological quantity.

The bulk-edge correspondence implies that an interface of the two topological platforms supports $\left|\delta\mathcal{C}_{\text{gap}}\right|$ unidirectional gapless edge states. A generic interface in the extended space does not have an obvious real-space geometric interpretation. The exceptions are the $x = const.$ interfaces, which correspond to standard real-space interfaces between the two waveguides. It is implicit that the waveguides cross-sections are identical when they are embedded in a 3D space.

Let us investigate the consequences of the bulk-edge correspondence for an interface $x = const.$, let us say $x = 0$. By definition, the edge states in the extended space must be localized near $x = 0$ and have a variation along the synthetic dimension (coordinate $n$) of the form $e^{inq_s}$ with $q_s$ the wave number of the edge state in the synthetic dimension: $\Psi = \left[\psi_n\left(x\right)\right]$ with $\psi_n\left(x\right) = \psi\left(x\right)e^{inq_s}$ ($n$=0,±1, ±2,…). Evidently, the edge states projection into real-space ($\psi_0\left(x\right) = \psi\left(x\right)$) corresponds to a wave trapped at the interface $x = 0$ of the two waveguides. A fixed $q_s$ in extended space corresponds to a spatial shift $x_0^{(i)} = N_i \dfrac{q_s}{2\pi} a$ in real-space. Thus, as $q_s$ varies from $0$ to $2\pi$ the internal structure of the *i*-th waveguide is displaced by $N_i$ cells. For some particular combinations of the shifts the $x = 0$ interface can support trapped (localized) states, typically designated as Tamm states [41]. The bulk-edge correspondence establishes that the number of gapless Tamm states branches in real-space is precisely $\left|\delta\mathcal{C}_{\text{gap}}\right|$, which is another key result of the paper.

### C. Numerical examples

To illustrate the developed ideas, we consider the case where the "waveguides" are 1D photonic crystals formed by stacked dielectric slabs (see Fig. 1(a*i*) for the geometry of a generic binary photonic crystal). All the materials are nonmagnetic ($\mu = \mu_0$). The band structure of a 1D photonic crystal can be calculated with standard methods [42]. We denote $Z_L(x_0,\omega)$ and $Z_R(x_0,\omega)$ as the Bloch impedances of the (unbounded) photonic crystal calculated at the plane $x = x_0$ when looking at the left or right, respectively (Fig. 1(a*i*)). The band diagram and the Bloch impedances are numerically evaluated as explained in Appendix B.

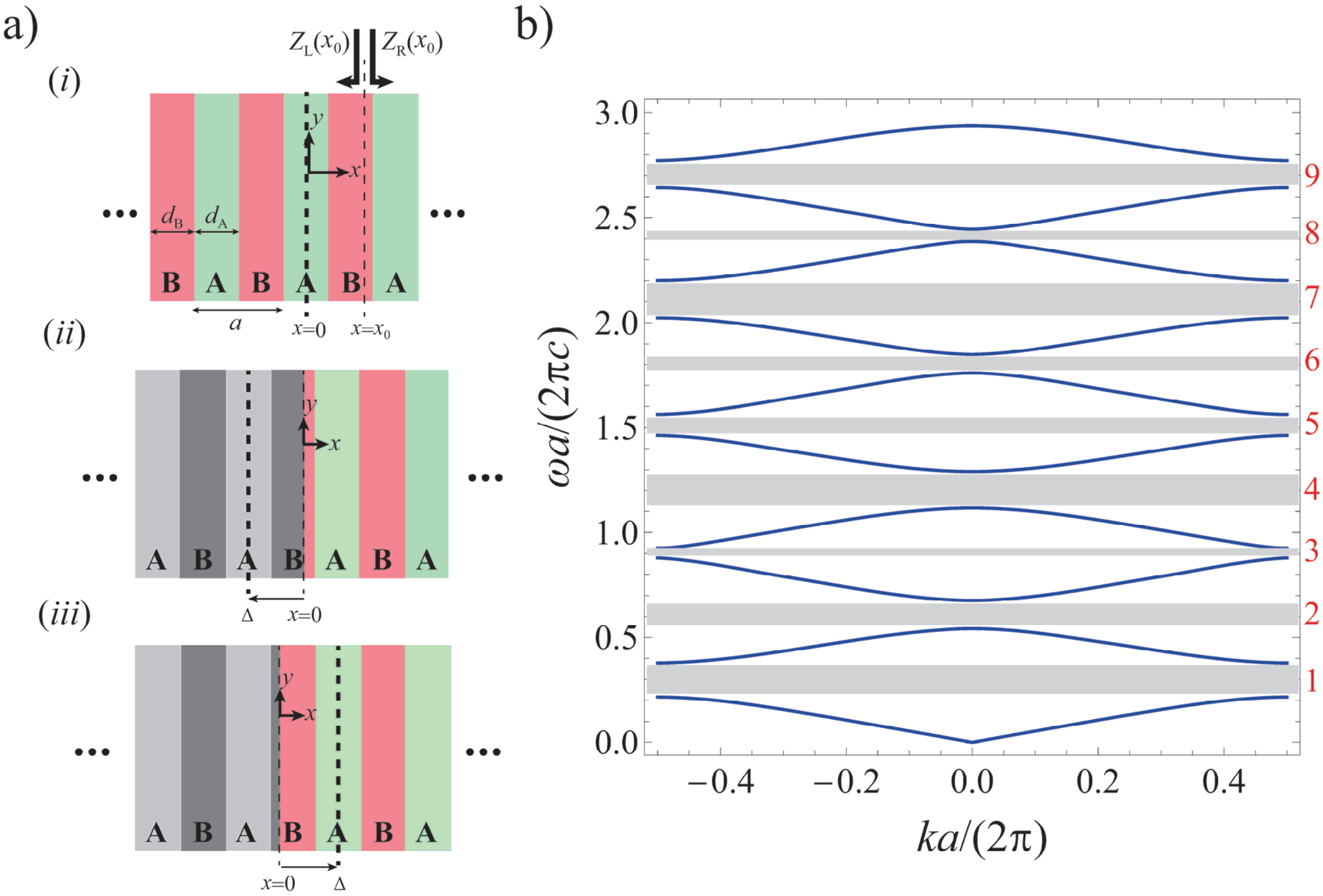

Fig. 1. (a) (*i*) Structure of a binary photonic crystal formed by two phases A and B. The left and right Bloch surface impedances calculated at the generic plane $x = x_0$ are indicated in the figure. The $x = 0$ plane is placed at the middle of slab A (center of symmetry). (*ii*) Representation of a negative displacement of the geometry of the photonic crystal. (*iii*) Representation of a positive displacement of the geometry. The areas shaded in grey are cut-way from the structure when another photonic crystal is inserted into the region $x<0$. (b) Band structure of a photonic crystal (blue solid curves) with parameters $\varepsilon_{\mathrm{A}}=7$, $\varepsilon_{\mathrm{B}}=1$, $d_{\mathrm{A}}=0.4a$ and $d_{\mathrm{B}}=0.6a$. The grey strips represent the band gaps. Each band gap is numbered with a red label.

Consider the scenario where two photonic crystals are paired to form an interface at $x=0$ (Fig. 2(a*i*)). The semi-space $x<0$ is filled with a photonic crystal modeled by $\hat{H}_1\left(x-x_0^{(1)}\left(q_s\right),-i\partial_x\right)$, and the semi-space $x>0$ by a photonic crystal modeled by $\hat{H}_2\left(x-x_0^{(2)}\left(q_s\right),-i\partial_x\right)$ with $x_0^{(i)}=N_i\Delta$. The trapped (defect-type) states at $x=0$ are the solutions of the characteristic equation [24]:

$$Z_L^{(1)}\left(-N_1\Delta,\omega\right)+Z_R^{(2)}\left(-N_2\Delta,\omega\right)=0. \tag{11}$$

Here, $Z_L^{(i)}$ and $Z_R^{(i)}$ are the left- and right- Bloch impedances of the *i*-th photonic crystal. Each value of $\Delta=\frac{q_s}{2\pi}a$ corresponds to a specific spatial-shift of the inner structure of the photonic crystals. In one full $q_s$-cycle, the parameter $\Delta$ varies from $\Delta=0$ to $\Delta=a$. The effect of shifting the geometry of a generic photonic crystal is illustrated in Figs. 1(a*ii*) and 1(a*iii*).

In the first example, we suppose that the semi-space $x<0$ is a perfectly electric conducting (PEC) wall, so that $Z_L^{(1)}=0$. The semi-space $x>0$ is filled with a binary photonic crystal with a unit cell formed by two dielectric slabs A and B of thickness $d_{\mathrm{A}}$ and $d_{\mathrm{B}}$, and relative dielectric permittivity $\varepsilon_{\mathrm{A}}$ and $\varepsilon_{\mathrm{B}}$, respectively [see Fig. 1(a)]. The structural parameters are taken as $\varepsilon_{\mathrm{A}}=7$, $\varepsilon_{\mathrm{B}}=1$, $d_{\mathrm{A}}=0.4a$ and $d_{\mathrm{B}}=0.6a$. Figure 1(b) shows the numerically calculated band structure ($\omega$ *vs* $k\equiv q_x$) with the band gaps shaded in grey. Since for the PEC semi-space $\mathcal{C}_{\mathrm{gap},1}=0$, it follows that the gap-Chern number difference is $\delta\mathcal{C}_{\text{gap}} = n_{\text{bands},2} \times N_2$ [Eq. (10)]. Interestingly, the gap Chern number of the dielectric photonic crystal with the synthetic dimension is nonzero, even though the structure in real-space is reciprocal. In typical systems, reciprocity (time reversal symmetry) implies a trivial Chern topology [4]. In contrast, in our problem the time-reversal symmetry in real-space does not imply the time-reversal symmetry in the extended space, i.e., it does not imply a trivial topology.

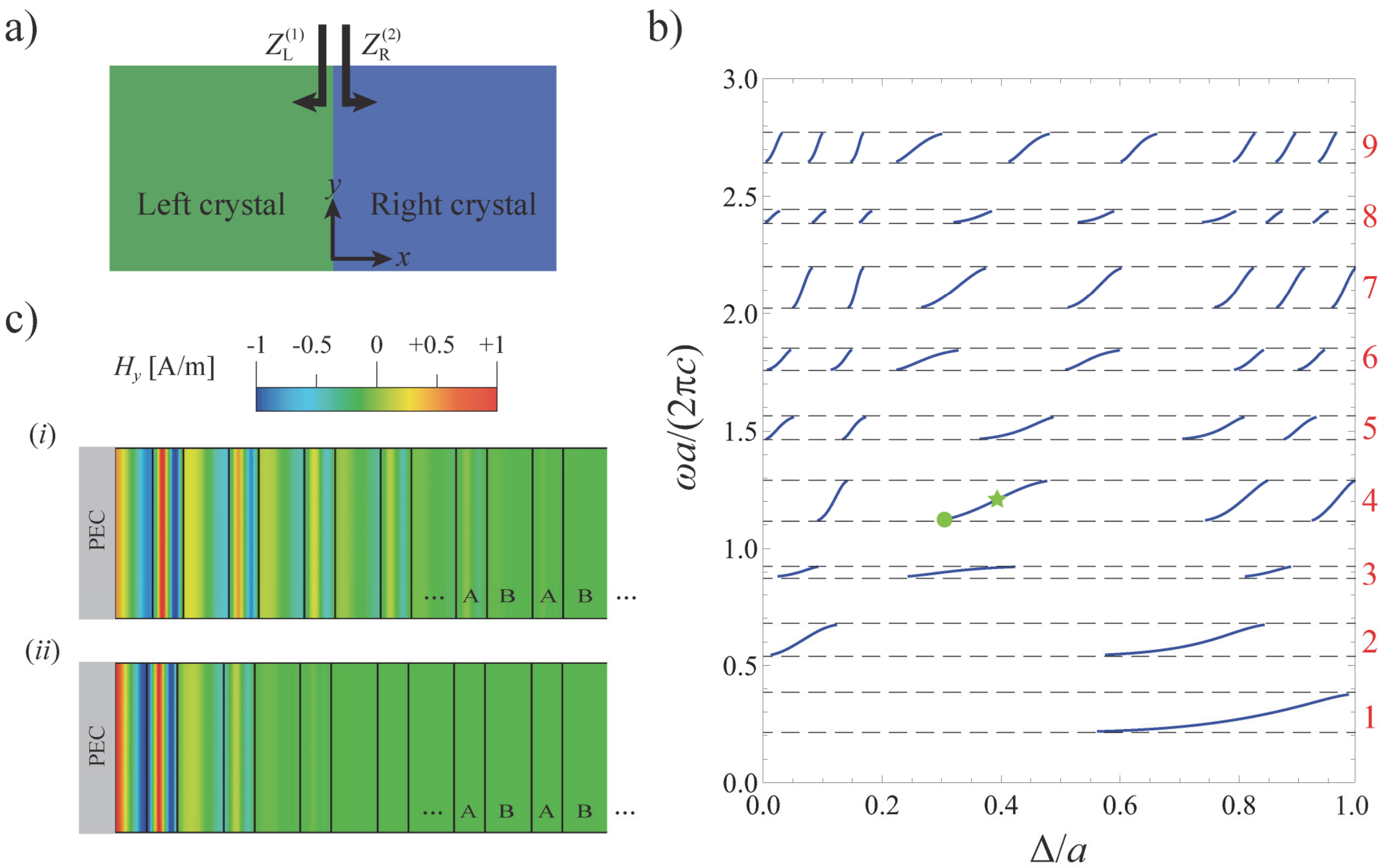

Fig. 2. (a) Representation of the pairing of two different photonic crystals. (b) Interface state solutions for a photonic crystal with the same parameters as in Fig. 1b in the right semi-space, and a PEC material in the left semi-space. The grey horizontal dashed lines delimit the band gaps which are numbered by the red labels. (c) Time-snapshot of the magnetic field of the interface state in the fourth gap for (*i*) the solution marked by the green circle with $\Delta/a = 0.3$, and (*ii*) the solution marked by the green star with $\Delta/a = 0.38$.

Suppose that $N_2 = -1$ so that the photonic crystal is displaced by a complete period to the negative $x$-axis in a full $\Delta$-cycle [Fig. 1(a*ii*)]. The dispersion of the interface states as a function of the spatial shift $\Delta$ is determined from $Z_R^{(2)}(\Delta,\omega) = 0$. The corresponding solutions in the band-gaps are plotted in Fig. 2(b) (blue curves). As seen, in agreement with the bulk-edge correspondence, $\delta\mathcal{C}_{\text{gap}} = -n_{\text{bands},2}$, the number of branches $\omega = \omega_n(\Delta)$ in each gap is exactly coincident with the number of bands below the gap. Each branch $\omega_n(\Delta)$ crosses completely the band-gap, and all the branches have a positive slope vs. $\Delta$ indicating that they are unidirectional gapless states in the extended space with the synthetic dimension. Our formalism enables to predict in a simple way how many (defect-type) trapped states occur in real-space for a fixed frequency in the band-gap, when the geometry of the crystal is displaced by one period. The number of Tamm state branches is exactly the number of bands below the gap, i.e., it equals the topological invariant. The operation of a "spatial shift" by one period may be regarded as a "topological pump" that inserts topological charge into the system, with the topological charge identical to the number of bands below the gap.

The profile of two trapped states in the fourth band gap are represented in Fig. 2(c). The field profiles were obtained using CST Studio Suite [43]. As seen, the trapped states are confined to the boundary of the photonic crystal, and decay exponentially into the bulk region. As could be expected, the trapped mode in the center of the band gap (Fig. 2(c)(*ii*) for $\Delta = 0.38$) is much more confined to the interface than the one near the bottom edge of the band gap (Fig. 2(c)(*i*) for $\Delta = 0.30$).

In the second example, the PEC region in the semi-space $x < 0$ is replaced by a binary photonic crystal with parameters $\varepsilon_{\text{A,l}} = 2$, $\varepsilon_{\text{B,l}} = 1$, $d_{\text{A,l}} = 0.4a$ and $d_{\text{B,l}} = 0.6a$, with the photonic crystal in the semi-space $x > 0$ the same as before. Figure 3(a) shows the band structures of the right (blue solid curve) and left (green dashed curve) crystals. There are two common frequency band gaps highlighted with the shaded grey strips. Consider first the situation wherein one of the photonic crystals is held fixed, while the other crystal is displaced by one cell period to the negative *x* direction.

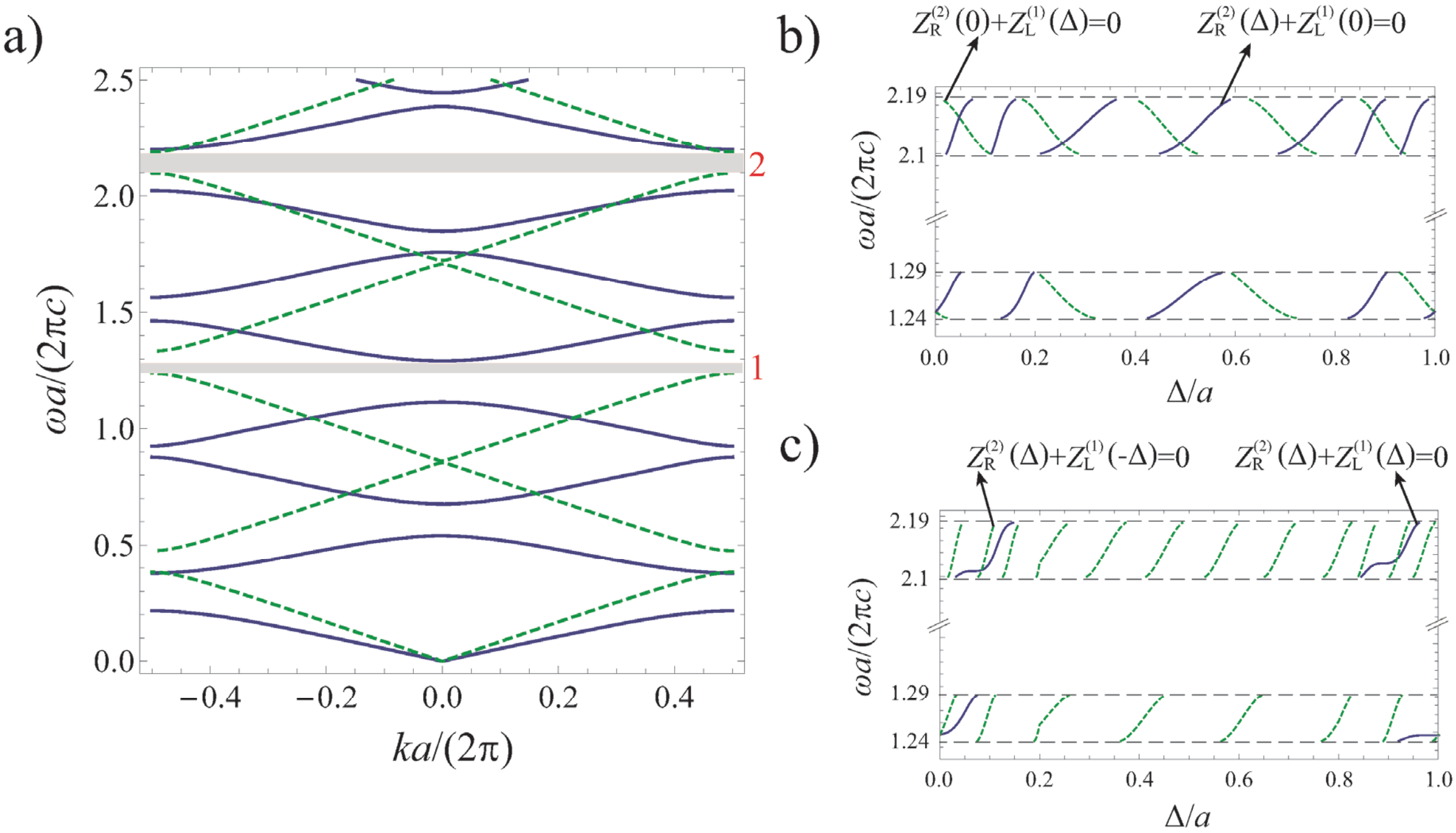

Fig. 3. (a) Green dashed curves: band structure of binary photonic crystal with parameters $\varepsilon_{A,1}=2$, $\varepsilon_{B,1}=1$, $d_{A,1}=0.4a$ and $d_{B,1}=0.6a$. Blue solid curves: band structure of the same photonic crystal as in Fig. 1. The grey strips indicate the common band gaps and the red labels the gap number. (b) Dispersion of the interface states in the common band gaps for a negative displacement of one of the crystals with the other held fixed. Blue solid curves: right photonic crystal slides one cell to the left; green dashed curves: left photonic crystal slides one cell to the left. (c) Similar to b) but for a situation where both crystals geometries suffer a negative spatial shift (blue solid curves, $N_1 = N_2 = -1$), or alternatively the left crystal suffers a positive spatial shift and the right crystal a negative spatial shift (green dashed curves, $N_1 = 1$ and $N_2 = -1$).

Figure 3(b) depicts the interface states dispersion $\omega = \omega_n(\Delta)$ in the two common gaps for the two possible displacements: i) the left crystal is held fixed and the right crystal slides to the left [blue solid curves; $N_1 = 0$ and $N_2 = -1$ in Eq. (11)], ii) the right crystal is held fixed and the left crystal slides to the left [green dashed curves; $N_1 = -1$ and $N_2 = 0$ in Eq. (11)]. For the case i) [case ii)] the number of solution branches is identical to the number of bands of the right [left] crystal below the gap, consistent with the bulk-edge correspondence [see Eq. (10)]. The slope of the curves $\omega = \omega_n(\Delta)$ is different in the two cases. This property is explained by the fact that $\delta\mathcal{C}_{\mathrm{gap}}$ has a different sign in each case. Indeed, the sign of $\delta\mathcal{C}_{\mathrm{gap}}$ is linked to the angular momentum of the edge modes in a closed system [38, 44, 45]. Thereby, the direction of the energy flow in the extended space must change when the gap Chern number sign changes.

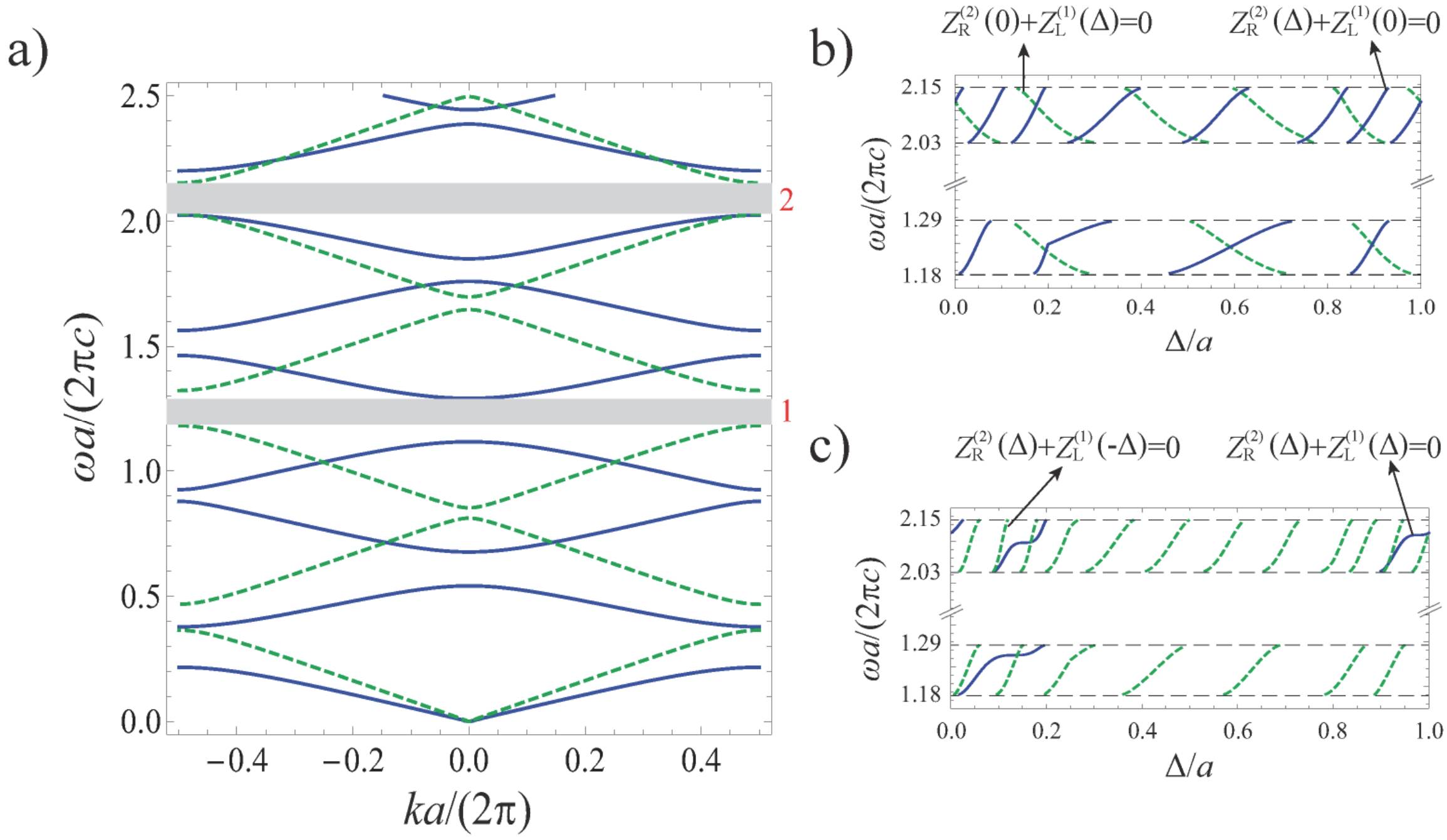

Fig. 4. Analogous to Fig. 3a for the case where the photonic crystal in the semi-space $x<0$ is replaced by a ternary photonic crystal with parameters $\varepsilon_{A,l}=2$, $\varepsilon_{B,l}=1$, $\varepsilon_{C,l}=3$, $d_{A,l}=0.3a$, $d_{B,l}=0.6a$ and $d_{C,l}=0.1a$. The band structure of the ternary photonic crystal is represented with green dashed curves in panel a).

We also studied the situations where the photonic crystals are simultaneously displaced to the negative *x*-direction $N_1=N_2=-1$ (blue solid curves in Fig. 3c), or, alternatively, the right crystal is displaced to the negative *x*-direction and the left crystal to the positive *x*-direction ($N_1=1$ and $N_2=-1$) (green dashed curves in Fig. 3c). In both cases, it is observed that the number of trapped states branches in a common band gap is identical to $\left|\delta\mathcal{C}_{\text{gap}}\right|\equiv\left|n_{\text{bands},2}\times N_2-n_{\text{bands},1}\times N_1\right|$. For example, for the lowest frequency gap $n_{\text{bands},2}=4$ and $n_{\text{bands},1}=3$. Consistent with this property there is a single gapless trapped state branch when $N_1=N_2=-1$ and 7 gapless trapped states branches when $N_1=-N_2=1$.

We verified that the bulk-edge correspondence also holds true for other more complex 1D photonic crystal geometries. For example, suppose that the left photonic crystal of the previous example is replaced by a ternary layered structure with parameters $\varepsilon_{\mathrm{A,l}}=2$, $\varepsilon_{\mathrm{B,l}}=1$, $\varepsilon_{\mathrm{C,l}}=3$, $d_{\mathrm{A,l}}=0.3a$, $d_{\mathrm{B,l}}=0.6a$ and $d_{\mathrm{C,l}}=0.1a$. Different from the binary crystals considered in the previous examples, the ternary crystal does not have inversion (parity) symmetry. Figure 4 reports a study identical to that of Fig. 3, when the ternary photonic crystal (region $x<0$) is paired with the same binary photonic crystal as in Fig. 2. The results are qualitatively analogous to those of Fig. 3 and again confirm that it is possible to predict the number of trapped states from the knowledge of the number of bands below the band-gap. Curiously, in this example the slope of the trapped states dispersion can be discontinuous (see the low-frequency gap in Fig. 4b, blue lines). This feature is due to the discontinuity of the permittivity profile of the photonic crystals.

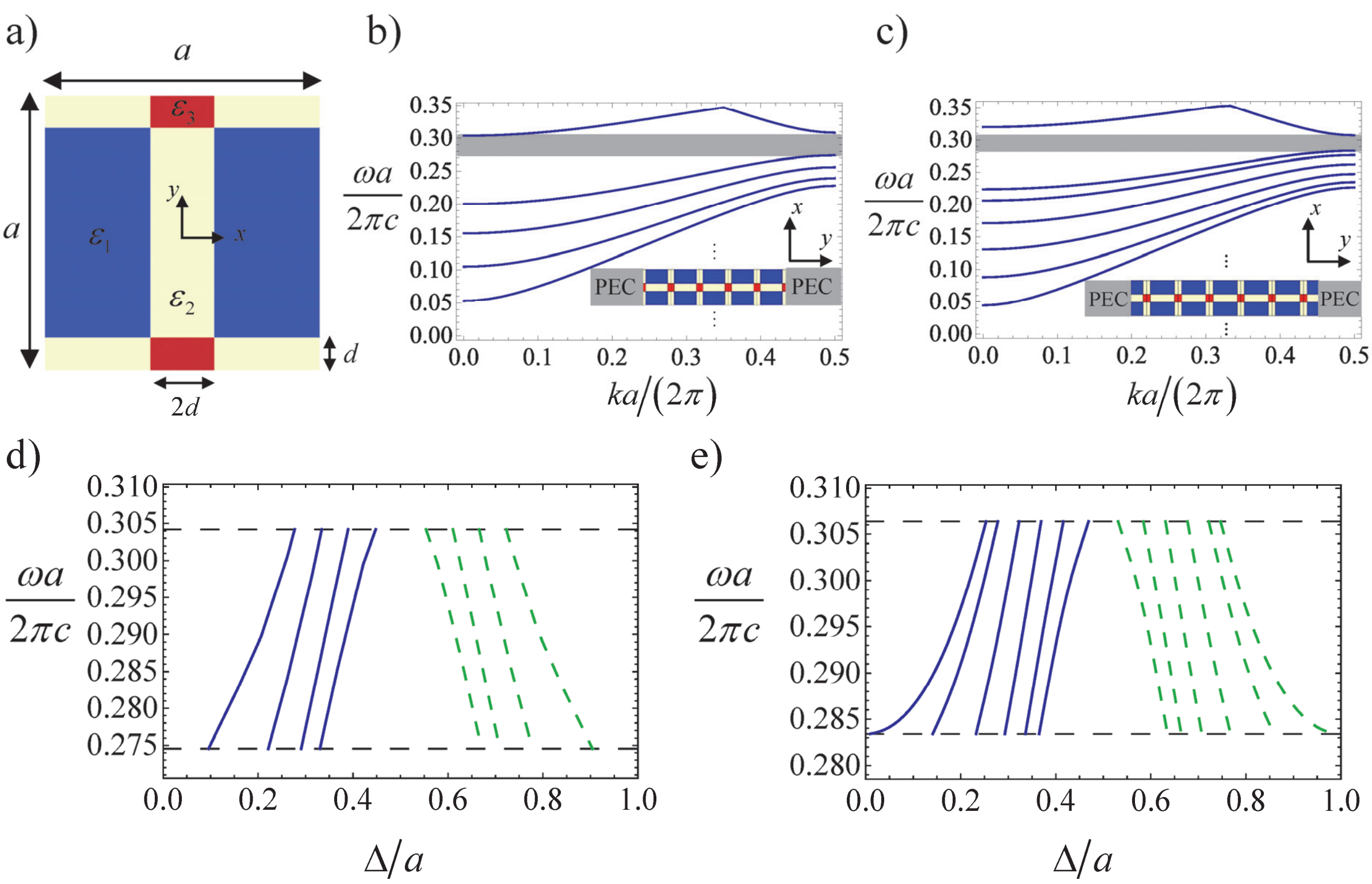

**Fig. 5.** a) Unit cell of a 2D photonic crystal with $\varepsilon_1=1$, $\varepsilon_2=6.5$, $\varepsilon_3=12$. The unit cell period is *a*. The parameter *d* represented in the figure is $d=0.115a$. b) Band diagram of the waveguide with metallic lateral walls constructed from a 2D photonic crystal with 5 unit cells along the *y*-direction. The shaded grey strip indicates the band gap of the waveguide. c) Similar to b) but for a waveguide constructed from a 2D photonic crystal with 6 unit cells along the *y*-direction. d) Dispersion of the interface states in the band gap for a waveguide constructed from a 2D photonic crystal with 5 unit cells

along the $y$-direction, placed in the semi space $x > 0$ and terminated with a metallic plate placed at $x = 0$. The waveguide geometry is continuously displaced by one lattice period along the $x$-direction. Blue solid curves: negative displacement ($N_2 = -1$). Green dashed curves: positive displacement ($N_2 = 1$). e) Similar to d) but for a waveguide constructed from a 2D photonic crystal with 6 unit cells along the $y$-direction.

Furthermore, we also studied the emergence of interface states in 1D-type waveguides embedded in a two-dimensional real-space. Specifically, consider a waveguide with metallic lateral walls constructed from a 2D photonic crystal with the unit cell represented in Fig. 5a. The lateral width of the guide is $N_y a$ and the electric field is oriented along the $z$-direction. The corresponding band-diagram for propagation along the $x$-direction is represented in Figs. 5b) and 5c) for the cases $N_y = 5$ and $N_y = 6$, respectively. The band diagram is numerically calculated with CST Studio Suite [43]. The band-gaps are shaded in gray. We terminated this waveguide (positioned in the semi-space $x > 0$) with a metallic plate placed at $x = 0$, and numerically found the edge states for different shifts of the waveguide geometry. The trapped states dispersion $\omega = \omega_n(\Delta)$ is shown in Figs. 5d) and 5e). We consider displacements along the negative ($N_2 = -1$, blue curves) and positive ($N_2 = 1$, green dashed curves) $x$-axis. As seen, also for this more complex system, the number of branches agrees with the number of bands of the waveguide below the gap. Furthermore, as expected, the slope of the curves $\omega = \omega_n(\Delta)$ depends on the displacement direction.

Finally, we present an example of a non-Hermitian photonic crystal. For simplicity the material dispersion is ignored here. The geometry of the photonic crystals is as in Fig. 1, except that the permittivity of the material A is taken equal to $\varepsilon_{\mathrm{A}} = 7 + 0.1i$, so that the photonic crystal is lossy. We calculated the complex band structure of the photonic crystal $\omega(q_x) = \omega' + i\omega''$ using the plane wave method [42]. Figure 6 shows a parametric plot of $\omega(q_x)$ in the complex plane with $-\pi / a < q_x < \pi / a$, i.e., the figure represents the projected band structure. As seen, the projected band structure is formed by disconnected regions (blue curves). Each disconnected region corresponds to a bulk band [13]. We determined the dispersion of the (complex) Tamm states when the photonic crystal is paired with a PEC wall. To this end, we solved $Z_R^{(2)}(\Delta,\omega)=0$ for $\Delta=\frac{q_s}{2\pi}a$ real-valued ($N_2=-1$). The green lines in Fig. 6 represent a parametric plot of the dispersion of the Tamm states $\omega(q_s)$ in the complex plane. In agreement with the bulk-edge correspondence, the number of Tamm state branches (in green) in each gap is identical to number of bulk bands (in blue) on the left of the gap. The Tamm states link the different bulk bands, and thereby are gapless. These properties confirm the topological nature of the Tamm states supported by non-Hermitian photonic crystals. Both the bulk bands and the edge states modes have an oscillation frequency such that $\omega''=\text{Im}\{\omega\}<0$ due to the effect of loss.

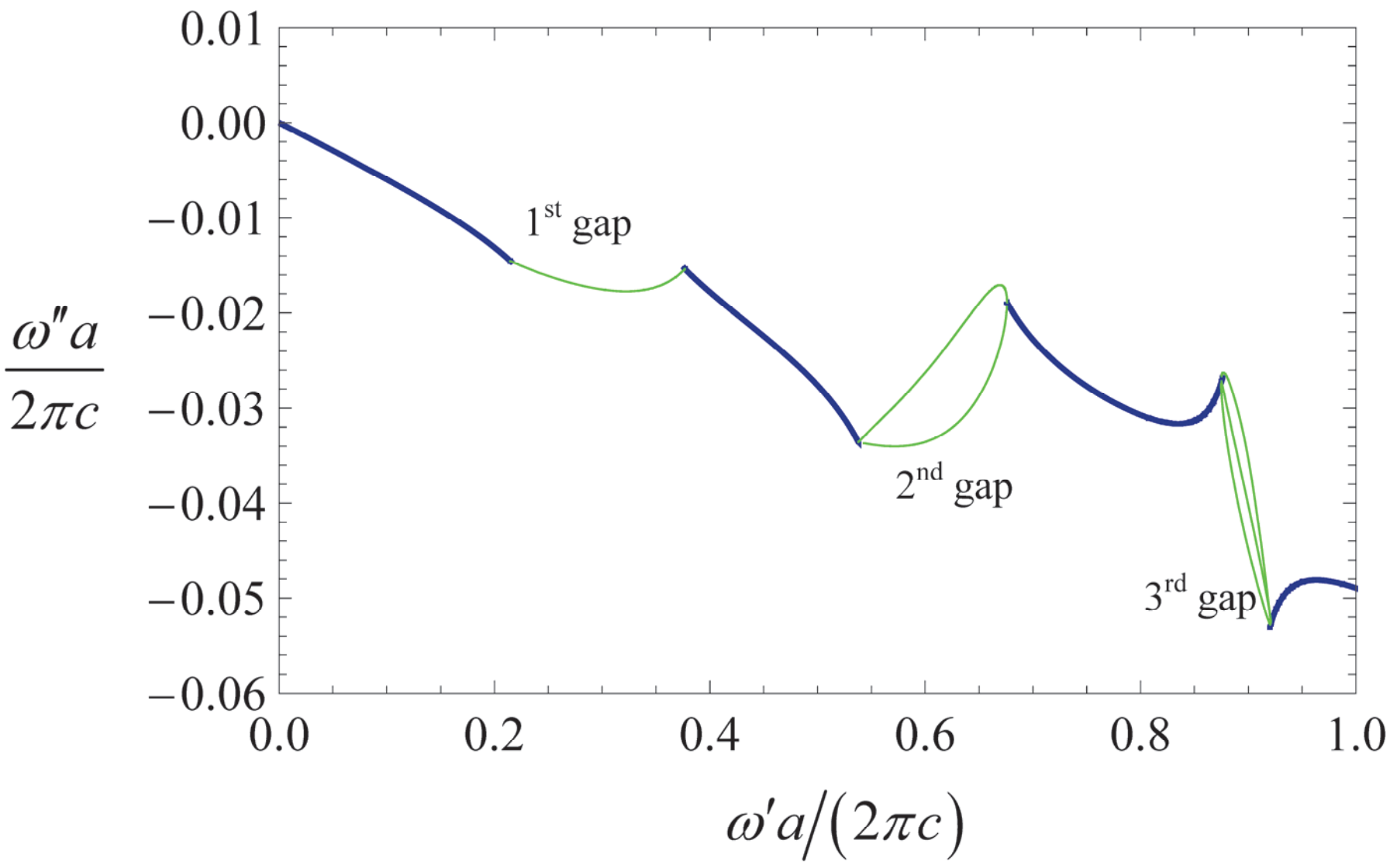

**Fig. 6.** Projected band structure $\omega=\omega'+i\omega''$ of a non-Hermitian photonic crystal with parameters $\varepsilon_A=7+0.1i$, $\varepsilon_B=1$, $d_A=0.4a$ and $d_B=0.6a$. The blue solid curves represent the different bulk bands. The green curves represent the projected dispersion of the edge states (Tamm states) of the extended synthetic system. The edge modes are gapless as they connect different bands. The number of Tamm states branches in a gap is identical to the gap number.

# IV. Conclusions

In summary, it was shown that in one-dimensional periodic systems the number of bands below a gap can be understood as a Chern topological number of an extended system with a synthetic dimension. This topological number determines the number of edge states in the extended space with the synthetic dimension. The real-space projection of the edge states are modes localized at the boundary of the 1D crystal (Tamm states), for some shift of the unit cell. The number of Tamm states branches "pumped" by a full-lattice period displacement equals the number of bands below the gap. This result is not rooted in any particular symmetry and is valid for non-Hermitian operators. Thereby, our work establishes a rigorous and simple bulk-edge correspondence for one-dimensional systems and uncovers a different topological mechanism to localize light at an interface of two arbitrary photonic waveguides.

**Acknowledgements:** This work was partially funded by the Institution of Engineering and Technology (IET), by the Simons Foundation and by Instituto de Telecomunicações under Project No. UID/EEA/50008/2020. Solange V. Silva acknowledges financial support by Fundação para a Ciência e a Tecnologia (FCT/POPH) and the cofinancing of Fundo Social Europeu under the PhD fellowship SFRH/BD/105625/2015. D. E. Fernandes acknowledges support by FCT, POCH, and the cofinancing of Fundo Social Europeu under the fellowship SFRH/BPD/116525/2016. T. A. Morgado acknowledges financial support by FCT under the CEEC Individual 2017 contract as assistant researcher with reference CT/Nº004/2019-F00069 established with IT – Coimbra.

## Appendix A: Calculation of the Chern number for a non-Hermitian system

In this Appendix, we extend the theory of Sect. II to non-Hermitian operators $\hat{H}$. In such a case the Berry connection must be defined using both the left and right eigenvectors of $\hat{H}$ [13, 17]. If the operator $\hat{H}$ is diagonalizable, it is possible to pick the left and right eigenvectors such that $\langle u^L_{q_x,n} | u^R_{q_x,m} \rangle = \delta_{nm}$, where the indices $n$ and $m$ identify different photonic bands. Here, $u^R_{q_x,m}, u^L_{q_x,n}$ are the envelopes of the Bloch eigenmodes ($\psi^R_{q_x,m}(x)$ and $\psi^{L}_{q_x,n}(x)$) of the operators $\hat{H}$ and $\hat{H}^{\dagger}$, respectively [13, 17]. Typically, $\hat{H}$ describes a system with material loss, whereas $\hat{H}^{\dagger}$ describes the corresponding time-reversal symmetric system with material gain. In the following, it is assumed that $\psi^{R}_{q_x,m}(x)$ and $\psi^{L}_{q_x,n}(x)$ are periodic functions of $q_x$.

We denote $\hat{H}^{R}_{\mathbf{q}} = \hat{H}_{\mathbf{q}}$ and $\hat{H}^{L}_{\mathbf{q}} = \hat{H}^{\dagger}_{\mathbf{q}}$ as the right- and left- Hamiltonians of the system with the synthetic dimension, obtained from $\hat{H}$ and $\hat{H}^{\dagger}$ using the same procedure as in the main text. Let us compute the Chern number associated with a generic (isolated) photonic band of the extended system. The Berry connection is defined by $\boldsymbol{\mathcal{A}}_{\mathbf{q}} = i\left\langle u^{L}_{\mathbf{q}} \mid \partial_{\mathbf{q}} u^{R}_{\mathbf{q}} \right\rangle$ (the band index is dropped here) [13, 17], with

$$u^{i}_{\mathbf{q}}(x) = \psi^{i}_{q_x}\left(x - x_0(q_s)\right) e^{-iq_x x} = u^{i}_{q_x}\left(x - x_0(q_s)\right) e^{-iq_x x_0(q_s)}. \qquad i=L,R \tag{A1}$$

and $\psi^{R}_{q_x}(x)$ and $\psi^{L}_{q_x}(x)$ the relevant Bloch eigenmodes of the 1D operators $\hat{H}$ and $\hat{H}^{\dagger}$, respectively. Using the same arguments as in the main text, it is readily seen that the Chern number is still given by Eq. (3). Furthermore, now the second component of the Berry potential is $\mathcal{A}_{2,\mathbf{q}} = i\left\langle \psi^{L}_{\mathbf{q}} \mid \partial_{q_s} \psi^{R}_{\mathbf{q}} \right\rangle$, which is evidently a periodic function of $q_x$. Hence, similar to the main text it is found that:

$$\mathcal{C} = \frac{1}{2\pi} \int_{-\pi/a}^{\pi/a} dq_x \left( \mathcal{A}_{1,\mathbf{q}}\Big|_{q_s=-\pi} - \mathcal{A}_{1,\mathbf{q}}\Big|_{q_s=+\pi} \right). \tag{A2}$$

Using $u^{i}_{\mathbf{q}}(x) = u^{i}_{q_x}\left(x - x_0(q_s)\right) e^{-iq_x x_0(q_s)}$ [Eq. (A1)] and $\left\langle u^{L}_{q_x} \mid u^{R}_{q_x} \right\rangle = 1$ one readily finds that $\mathcal{A}_{1,\mathbf{q}} = i\left\langle u^{L}_{q_x}\left(x - x_0(q_s)\right) \mid \partial_{q_x}\left[u^{R}_{q_x}\left(x - x_0(q_s)\right)\right]\right\rangle + x_0(q_s)$. The first term of $\mathcal{A}_{1,\mathbf{q}}$ is a periodic function of the synthetic momentum $q_s$, and thereby the result of the main text [Eq.(5)] remains valid in the non-Hermitian case.

## Appendix B: Dispersion equation and Bloch impedance of a 1D photonic crystal

Here, we derive the characteristic equation for the Bloch waves of a 1D layered photonic crystal and the Bloch wave impedances. The unit cell is formed by an arbitrary number ($N$) of layers (see Fig. 7 for the case $N$=3).

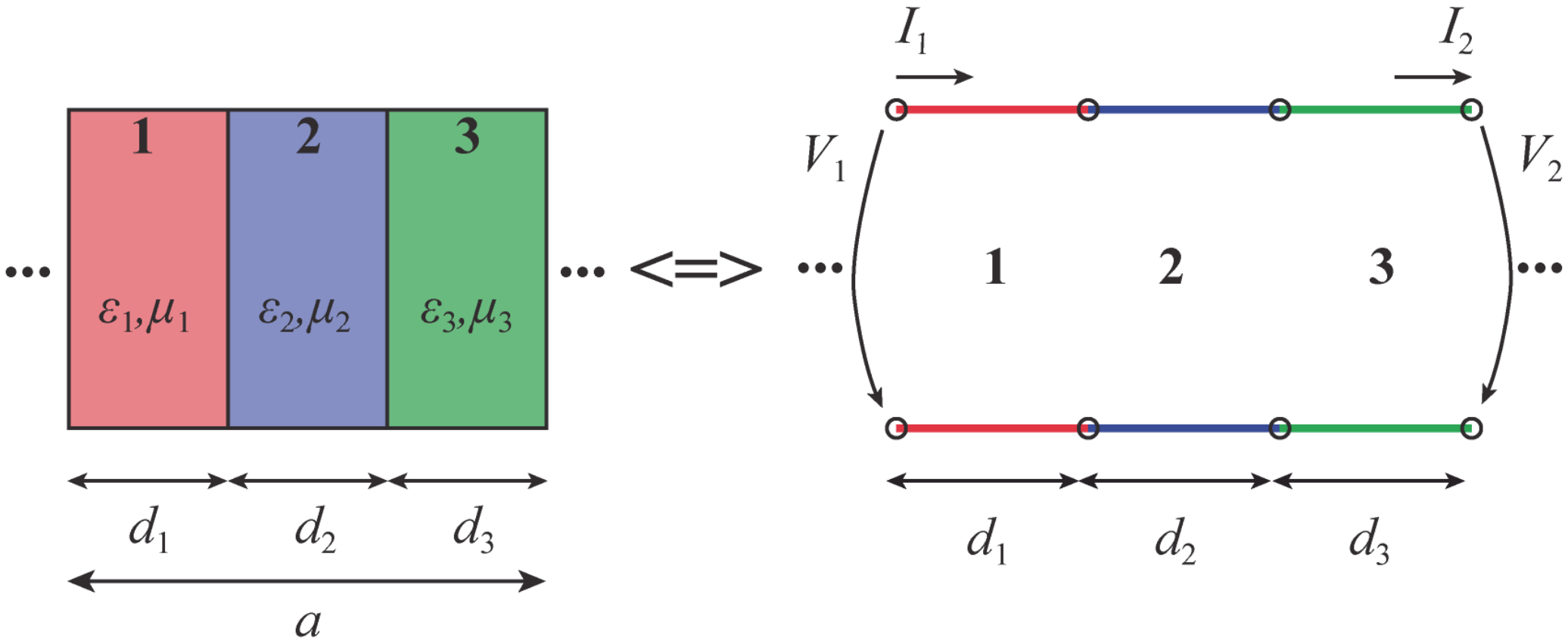

**Fig. 7**. Equivalence between a multi-layered photonic crystal and a periodic transmission line.

As is well-known, the wave propagation in a 1D photonic crystal is formally equivalent to the propagation in a periodic transmission line (Fig. 7). Thus, the characteristic equation for the Bloch waves can be easily found using the ABCD-matrix formalism [46]. To this end, one needs to find the ABCD-matrix for a unit cell, which links the input and output voltages and currents as:

$$\begin{pmatrix} V_1 \\ I_1 \end{pmatrix} = \begin{pmatrix} A & B \\ C & D \end{pmatrix}_{\text{global}} \begin{pmatrix} V_2 \\ I_2 \end{pmatrix}. \tag{B1}$$

From the theory of microwave networks, the global ABCD matrix is given by the product of the ABCD matrices of the uniform line sections:

$$\mathbf{M} \equiv \begin{pmatrix} A & B \\ C & D \end{pmatrix}_{\text{global}} = \begin{pmatrix} A & B \\ C & D \end{pmatrix}_1 \cdots \begin{pmatrix} A & B \\ C & D \end{pmatrix}_N . \tag{B2}$$

In the above,

$$\begin{pmatrix} A & B \\ C & D \end{pmatrix}_i = \begin{pmatrix} \cosh(\gamma_i d_i) & Z_{c,i} \sinh(\gamma_i d_i) \\ Z_{c,i}^{-1} \sinh(\gamma_i d_i) & \cosh(\gamma_i d_i) \end{pmatrix}, \qquad i = 1,2,...,N \tag{B3}$$

is the ABCD matrix of the $i$th section, $Z_{c,i} = \eta_0 \sqrt{\mu_i / \varepsilon_i}$ is the wave impedance and $\gamma_i = -i \frac{\omega}{c} \sqrt{\mu_i \varepsilon_i}$ is the propagation constant. Here, $\eta_0$ is the free-space impedance and $c$ is the speed of light in vacuum. The permittivity and permeability $\varepsilon_i, \mu_i$ are normalized to the free-space values.

For Bloch waves the input and output voltages are linked by $\begin{pmatrix} V_2 \\ I_2 \end{pmatrix} = e^{-\gamma a} \begin{pmatrix} V_1 \\ I_1 \end{pmatrix}$, with $\gamma = \alpha - ik$ the (complex) propagation constant of the Bloch mode. Thus, using Eq. (B1) one finds that the "output" voltages and currents satisfy the homogeneous equation:

$$\left(\mathbf{M} - \mathbf{1} e^{+\gamma a}\right) \cdot \begin{pmatrix} V_2 \\ I_2 \end{pmatrix} = 0, \tag{B4}$$

with $\mathbf{M}$ the global ABCD matrix defined as in Eq. (B2). This result implies that $\det(\mathbf{M} - \mathbf{1}\lambda) = 0$, or equivalently $\lambda^2 - \lambda \operatorname{tr}(\mathbf{M}) + \det(\mathbf{M}) = 0$, with $\lambda = e^{+\gamma a}$. Since the system under analysis is reciprocal, one has $\det(\mathbf{M}) = 1$ [46]. The solutions of the second degree equation are $\lambda_{1,2} = \frac{\operatorname{tr}(\mathbf{M})}{2} \pm \sqrt{\left[\frac{\operatorname{tr}(\mathbf{M})}{2}\right]^2 - 1}$. Because of $\lambda_{1,2} = e^{\pm \gamma a}$, one has $e^{+\gamma a} + e^{-\gamma a} = \lambda_1 + \lambda_2 = \operatorname{tr}(\mathbf{M})$. This implies that the characteristic equation for the Bloch waves is

$$\cosh(\gamma a) = \frac{\operatorname{tr}(\mathbf{M})}{2}. \tag{B5}$$

The photonic band structure of a Hermitian crystal is found by looking for solutions of the above equation with $\gamma = -ik$ a purely imaginary number.

Next, we derive the formulas of the Bloch impedances $Z_L(x_0,\omega)$ and $Z_R(x_0,\omega)$. Let us suppose without loss of generality that $x = x_0$ lies in the first line section, as illustrated in Fig. 8.

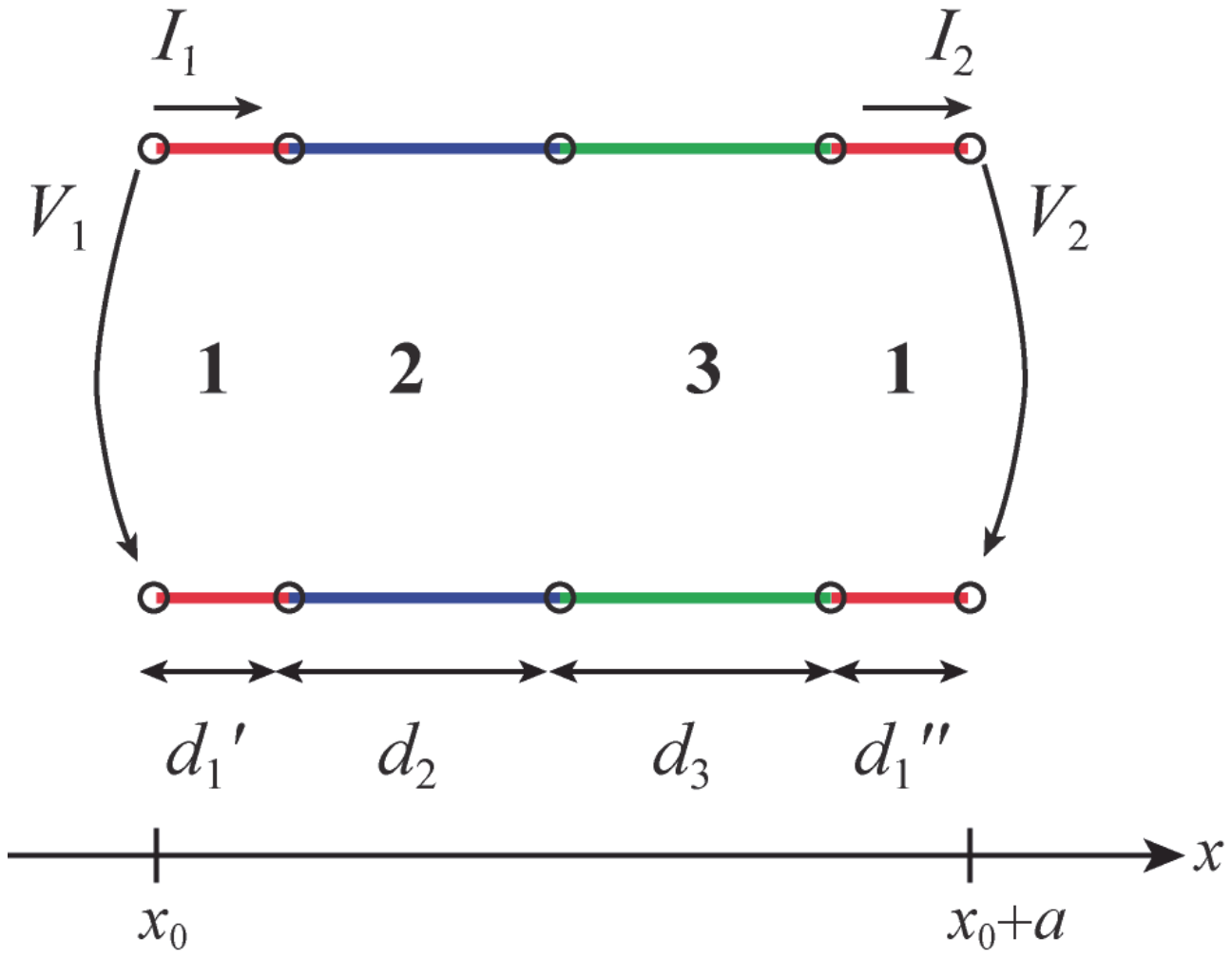

**Fig. 8**. Geometry used in the calculation of the Bloch impedance.

It is useful to obtain the global ABCD matrix ($\mathbf{M}(x_0)$) for one period, with the input and output voltages and currents referred to the planes $x = x_0$ and $x = x_0 + a$, respectively (Fig. 8). This is done as before by multiplying the ABCD matrices of the uniform line sections. For the example, for the geometry shown in Fig. 8 one has:

$$\mathbf{M}(x_0) = \begin{pmatrix} A & B \\ C & D \end{pmatrix}_{1,\, d_1'} \begin{pmatrix} A & B \\ C & D \end{pmatrix}_{2,d_2} \begin{pmatrix} A & B \\ C & D \end{pmatrix}_{3,\, d_3} \begin{pmatrix} A & B \\ C & D \end{pmatrix}_{1,d_1''}. \tag{B6}$$

Note that $d_1'$ and $d_1''$ depend on $x_0$. Similar to Eq. (B4), for Bloch waves associated with a propagation factor $e^{-(\pm\gamma)x}$ the output voltage and current satisfy:

$$\left(\mathbf{M}(x_0) - \mathbf{1}e^{\pm\gamma a}\right) \cdot \begin{pmatrix} V_2 \\ I_2 \end{pmatrix} = 0, \tag{B7}$$

Note that $\gamma = \gamma(\omega)$ depends exclusively on the frequency and can be found from Eq. (B5). In the band-gaps $\gamma$ is complex-valued and it is implicit that $\mathrm{Re}\{\gamma\} > 0$. Denoting $\mathbf{M}(x_0) = \begin{pmatrix} A & B \\ C & D \end{pmatrix}$, it follows from Eq. (B7) that:

$$\begin{pmatrix} V_2 \\ I_2 \end{pmatrix} \sim \begin{pmatrix} -B \\ A - e^{\pm\gamma a} \end{pmatrix} \sim \begin{pmatrix} D - e^{\pm\gamma a} \\ -C \end{pmatrix}. \tag{B8}$$

Hence, the Bloch impedance for a wave that propagates towards the positive *x*-direction is

$$Z_R(x_0) = \frac{V_2}{I_2} = \frac{-B}{A - e^{\gamma a}} = \frac{D - e^{\gamma a}}{-C}, \tag{B9a}$$

whereas the Bloch impedance for a wave that propagates towards the negative *x*-direction is:

$$Z_L(x_0) = \frac{V_2}{-I_2} = \frac{B}{A - e^{-\gamma a}} = \frac{D - e^{-\gamma a}}{C}. \tag{B9b}$$

We used the fact that the Bloch impedances are periodic: $Z_R(x_0) = Z_R(x_0 + a)$, etc.

The trapped states at an interface ($x = 0$) between two semi-infinite photonic crystals are forcibly decaying in space Bloch modes of the infinite photonic crystals in the regions $x > 0$ and $x < 0$. Thereby, since the current and voltage are continuous at the interface $x = 0$, the localized trapped states must satisfy $Z_L^{(1)}(x = 0, \omega) + Z_R^{(2)}(x = 0, \omega) = 0$, consistent with the main text.